\documentclass[aps,prx,twocolumn,showpacs,amsmath,amssymb]{revtex4-1}

\usepackage{graphicx}
\usepackage{color}
\usepackage{xcolor}
\usepackage{epstopdf}
\usepackage{amsmath}
\usepackage{bm}
\usepackage{hyperref}
\usepackage{cleveref}
\usepackage{tcolorbox}
\usepackage{braket}
\usepackage[export]{adjustbox}
\usepackage[hang]{subfigure}
\usepackage[T1]{fontenc}
\UseRawInputEncoding
\usepackage{float}
\usepackage{latexsym}

\begin{document}
	\title{Nonreciprocal and long-range three-body interactions in Bose-Einstein condensates induced by optical feedback}
	
	\author{Yi-Qing Zhang$^1$}
	\thanks{These authors contribute equally to this work.}
	
	\author{Liang-Jun He$^1$}
	\thanks{These authors contribute equally to this work.}
		
	\author{Han Pu$^2$}
	\email{hpu@rice.edu}
	
	\author{Zheng-Wei Zhou$^{3,4,5,6}$}
	\email{zwzhou@ustc.edu.cn}
		
	\author{Yong-Chang Zhang$^1$}
	\email{zhanyc@xjtu.edu.cn}
	\affiliation{$^1$MOE Key Laboratory for Nonequilibrium Synthesis and Modulation of Condensed Matter, and Shaanxi Key Laboratory of Quantum Information and Quantum Optoelectronic Devices, School of Physics, Xi'an Jiaotong University, Xi'an 710049, People's Republic of China\\
	$^2$Department of Physics and Astronomy, and Smalley-Curl Institute, Rice University, Houston, Texas 77005, USA\\
	$^3$Anhui Province Key Laboratory of Quantum Network, University of Science and Technology of China, Hefei, 230026, People's Republic of China\\
	$^4$Anhui Center for Fundamental Sciences in Theoretical Physics, University of Science and Technology of China, Hefei, 230026, People's Republic of China\\
	$^5$Synergetic Innovation Center of Quantum Information and Quantum Physics, University of Science and Technology of China, Hefei, 230026, People's Republic of China\\
	$^6$Hefei National Laboratory, University of Science and Technology of China, Hefei, 230088, People's Republic of China
    }

	\date{\today}
	
	\begin{abstract}

	We propose generating long-range and nonreciprocal three-body interactions in quantum gases via optical feedback. By placing a quasi-two-dimensional Bose-Einstein condensate (BEC) in front of two reflecting mirrors and illuminating it with dichromatic laser beams, these driving optical fields traverse the BEC twice, thereby inducing a feedback effect on the atoms. We demonstrate that this optical feedback gives rise to an effective three-body atom-atom interaction with remarkable long-range and nonreciprocal properties. Due to its long-range nature, this three-body interaction can cause unique spatial symmetry-breaking behaviors in the BEC, resulting in various stable stationary states as well as unexpected diffusive collapse. Notably, a distinct ring state emerges through a purely self-organizing process. Furthermore, by analyzing the real-time dynamics of the BEC, we show that the nonreciprocal nature of this interaction can lead to intriguing self-acceleration of the condensate, seemingly violating Newton's law of motion. Additionally, our scheme offers a highly controllable setting, where pairwise two-body interactions can be tuned to vanish. This flexibility provides a promising route for exploring exotic physics associated with multi-body interactions.
	
	\end{abstract}

	\maketitle

\section{Introduction}
\label{sec:1}

Particle-particle interactions are fundamental to many-body physics, playing a pivotal role in determining the properties of physical systems ranging from microscopic atomic ensembles to macroscopic galaxies. At the basic level, particles interact through pairwise two-body interactions, which are typically classified into two categories: contact interactions and long-range interactions depending on how the range of the interaction compares to the average distance between particles. In typical cold atom systems, interactions can be regarded as contact and are responsible for evaporative cooling to condense atomic ensembles~\cite{Cornell1995Sci,PhysRevLett.75.3969,RevModPhys.74.1131}, a crucial step in achieving Bose-Einstein condensates (BECs). These interactions are also responsible for various many-body phenomena, such as the collapse and expansion dynamics~\cite{Wieman2001Nature}, superfluid-Mott insulator phase transition~\cite{Bloch2002Nature}, and solitons induced by attractive forces~\cite{Salomon2002Sci,Hulet2002Nature,Wieman2006PRL,Robins2014PRL,Hullet2017Sci,Torner2019review}. The scale of long-range interactions, where particles exert influence on each other even at finite separations, leads to a wealth of additional many-body phenomena, including exotic states of matter. One notable example is the long-sought supersolid state~\cite{Chester:PRA:1970,Leggett:PRL:1970,Balibar2010review,Boninsegni2012RMP}. These states have been theoretically proposed to arise in systems such as Rydberg atoms~\cite{henkel2010three,PhysRevLett.108.265301,PhysRevA.99.063625,Cinti2014NC,Han2018PRL} and dipolar quantum gases~\cite{Lewenstein2002PRL,Lewenstein2003PRL,Zhang2019PRL}, which are governed by van der Waals and dipole-dipole interactions, respectively. In particular, supersolidity in dipolar BECs has been experimentally confirmed in recent years~\cite{Tanzi2019PRL,Chomaz2019PRX,Bottcher2019PRX,norcia2021two,Biagioni2022PRX}.

In quantum gases, while contact interactions arising from $s$-wave scattering can be easily tuned using the standard technique of Feshbach resonance, manipulating the inherent long-range interactions remains a challenge. However, leveraging the high flexibility of light-atom coupling systems, several schemes have been proposed to synthetically generate light-induced long-range interactions between atoms. These interactions offer greater controllability and exhibit unique characteristics compared to their inherent counterparts. For example, a superradiance-induced infinite long-range interaction has been realized in a cavity system, leading to supersolid-like states~\cite{Landig2016,Leonard2017Nature} and enabling metastable phase transitions~\cite{Hruby2018PNAS}. Similarly, photon-mediated long-range interactions have been proposed in systems such as strong coupled optical fields and BECs~\cite{Qin2015PRL,Qin2016PRA}, ring cavities~\cite{Helmut2018PRL,Helmut2021Review} and single-mirror feedback setups~\cite{PhysRevLett.121.073604,PhysRevA.103.023308,Robb2023PRR}. These interactions result in BECs that exhibit self-organized profiles, often accompanied by distinctive symmetry-breaking phenomena.

\begin{figure}[!t]
	\centering
	\includegraphics[width=\linewidth]{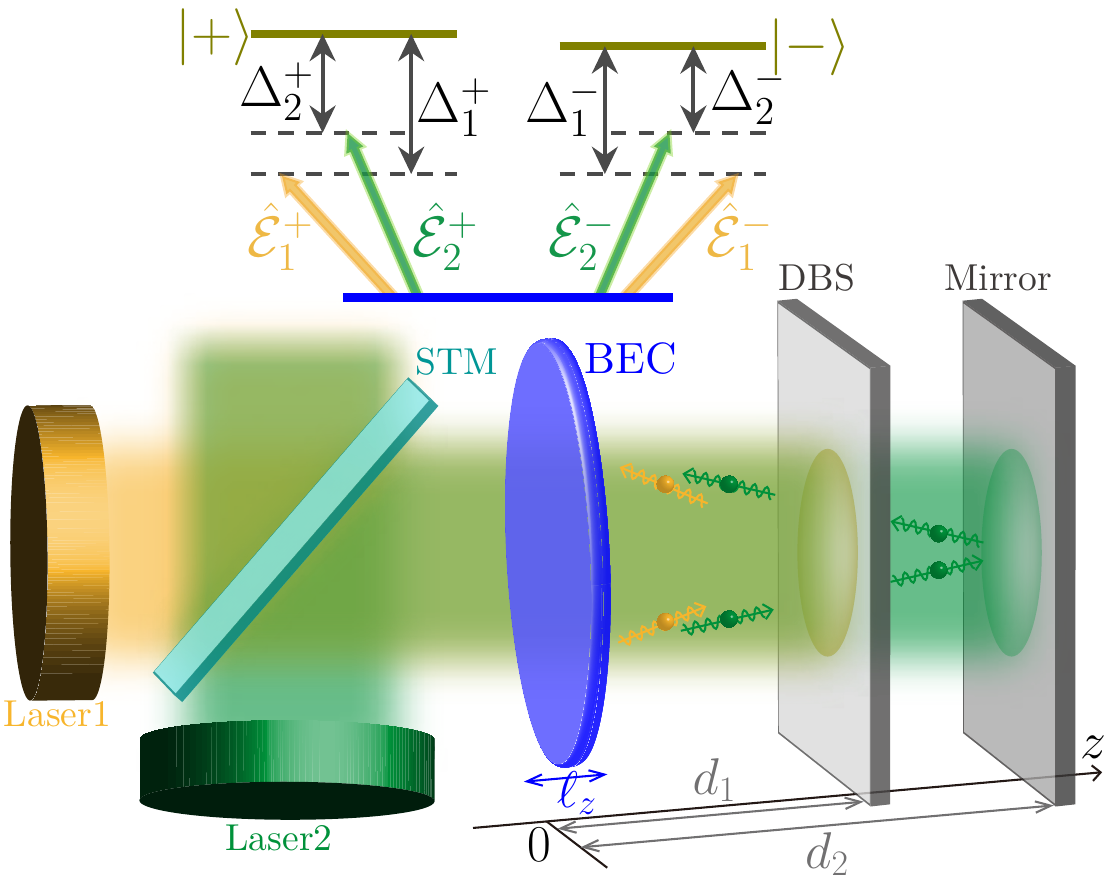}
	\caption{A schematic illustration of the proposed setup: A quasi-2D BEC with a thickness of $l_z$ is placed in front of a dichromatic beam splitter (DBS) and a normal reflecting mirror. The separations between the BEC and the two mirrors are $d_1$ and $d_2$, respectively. Two laser beams with different frequencies shine on the BEC through a semi-transparent mirror (STM), propagate towards and back-reflect from the DBS (yellow beam) and the normal mirror (green beam), respectively. The forward and backward lights interact with the atoms via a V-type coupling.}
	\label{fig1}
\end{figure}

Although two-body interactions are ubiquitous and form the foundation of many-body systems, higher-order interactions, such as three-body interactions, can lead to distinct phenomena, including nucleon clusters~\cite{RevModPhys.39.745}, topological orders~\cite{PhysRevA.82.052320,Kay2005}, few-body bound states~\cite{Hulet2009Sci,Yudkin2024NC}, collapse arrest~\cite{PhysRevLett.115.075303}, dynamical instability~\cite{Liu2019book}, and novel states of matter~\cite{PhysRevB.79.020503,PhysRevLett.101.150405,PhysRevB.83.184513,Bonnes_2010}. Due to their profound effects, three-body interactions have garnered considerable attention in recent years~\cite{PhysRevA.101.063610,PhysRevA.106.012435,Nath2022,PhysRevLett.122.103001,PhysRevLett.128.160505,10.1063/1.4986291,PhysRevA.108.043303,PhysRevD.97.014504,PhysRevA.97.013609,Abdullaev2015,PhysRevLett.103.140501,PhysRevA.93.033644}. However, studying three-body effects presents several challenges. On the one hand, three-body interactions are generally weaker than two-body interactions in most systems, making their effects difficult to observe, as they are often overshadowed by the stronger two-body effects~\cite{PhysRevE.74.021202,akhmediev1999bose,Gammal2000,Marcelli1999}. On the other hand, in cold-atom systems, three-body effects manifest primarily as three-body losses rather than elastic interactions~\cite{Kim2004PRA,Daley2009PRL,Schemmer2018PRL}, which merely reduce the lifetime of quantum gases. To address these challenges, various theoretical schemes have been proposed to generate effective three-body interactions in different systems, such as superconducting circuits~\cite{PhysRevA.101.062312}, optical lattices~\cite{Büchler2007,PhysRevLett.93.056402,PhysRevA.70.053620,Johnson_2009,PhysRevA.82.043629,PhysRevA.78.043603}, optical tweezers arrays~\cite{PhysRevLett.124.043402}, binary BECs or mixtures~\cite{PhysRevLett.112.103201,PhysRevLett.128.083401,Tajima2024PRA}, trapped ions~\cite{PhysRevA.79.060303,Andrade_2022,PhysRevLett.125.133602}, optical fields coupled with Rydberg atoms~\cite{PhysRevLett.117.113601,Bai2016OE}, and quantum dots~\cite{PhysRevA.81.052331}. Recently, three- and multi-body effects have been experimentally observed in, e.g., cold atoms in optical lattices~\cite{Pan4body2017,Goban2018}, photons~\cite{Liang2018}, Bose-Fermi mixtures~\cite{Chin2019Nature}, ion spins~\cite{Katz2023}, and lithium-7 atoms~\cite{Yudkin2024NC}. However, most previous research on multi-body interactions has been carried out in discrete lattice models~\cite{Büchler2007,PhysRevLett.93.056402,PhysRevA.70.053620,Johnson_2009,PhysRevA.82.043629,PhysRevA.78.043603,PhysRevLett.124.043402,PhysRevA.79.060303,Andrade_2022,PhysRevLett.125.133602,PhysRevA.81.052331,Katz2023}, while proposals for generating effective three-body interactions in continuum quantum gases are rare (see, however, Refs.~\cite{PhysRevLett.112.103201,PhysRevLett.128.083401}). Furthermore, in most previous schemes, three-body interactions appear as weaker contributions of higher order alongside the more dominant two-body interactions~\cite{PhysRevLett.93.056402,PhysRevA.70.053620,Johnson_2009,PhysRevA.82.043629,PhysRevA.78.043603,PhysRevLett.125.133602,Bai2016OE,Tajima2024PRA}, making it difficult to isolate and identify pure three-body effects.

Thus, exploring novel approaches to realize controllable multi-body interactions remains an important task. To this end, we propose here a new scheme to generate effective three-body interactions in a simple optical feedback system, consisting of a quasi-2D BEC and two reflecting mirrors, as shown in Fig.~\ref{fig1}. By pumping the system with dichromatic optical fields, we find that the optical feedback, coupled with diffraction effects, leads to the emergence of an effective three-body interaction between the atoms. In contrast to previous methods, our approach makes it possible to completely suppress the two-body contributions, offering a flexible platform for exploring the rich physics driven solely by three-body interactions. Moreover, the three-body interactions in our scheme exhibit unique long-range and nonreciprocal properties. To the best of our knowledge, such long-range and nonreciprocal three-body interactions have not been observed in any other system.

As a result of its long-range nature, the optical feedback-induced three-body interaction can lead to significant spatial symmetry breaking, giving rise to various density distribution patterns. More surprisingly, due to the nonreciprocal nature of the interaction, the BEC may exhibit self-acceleration behavior, which manifestly violates Newton's law of motion. While Newton's law of motion is a cornerstone of classical mechanics, it can be broken in systems with nonreciprocal interactions, where the action-reaction symmetry is disrupted. Such counterintuitive phenomena have attracted significant attention in recent years. Nonreciprocal interactions can arise in various contexts, such as active matters~\cite{Kryuchkov2018,PhysRevX.12.010501}, particles embedded in non-equilibrium environments~\cite{PhysRevX.5.011035,Chiu2023,Wu2023pnas}, particles with negative effective mass~\cite{Wimmer2013}, stroboscopic nonlinearity~\cite{Ma2023}, and soft matter systems~\cite{PhysRevLett.128.048002}. Although much of the existing work has focused on two-body non-reciprocal interactions, our proposal introduces the first example of a three-body nonreciprocal interaction in cold atom systems. This opens up new avenues for exploring novel quantum many-body physics related to nonreciprocal interactions.

The remainder of this paper is organized as follows. In Sec.~\ref{sec:2}, we introduce the physical setting as well as the derivation of three-body interaction. In Sec.~\ref{sec:3}, we discuss the stable stationary states led by this three-body interaction. In Sec.~\ref{sec:4}, we present the real-time dynamics of the BEC starting from symmetric and asymmetric initial states, respectively. We conclude in Sec.~\ref{sec:5}.

\section{Light-induced three-body interaction}
\label{sec:2}
As shown in Fig.~\ref{fig1}, we consider a quasi-two-dimensional (2D) BEC placed in front of two mirrors. The distances between the BEC and the two mirrors are $d_1$ and $d_2$, respectively. Two laser beams with frequencies $\omega_1$ and $\omega_2$ are directed onto the BEC. The first mirror is a dichromatic beam splitter (DBS) that selectively transmits the light of frequency $\omega_2$ while reflecting the other beam of frequency $\omega_1$~\cite{PhysRevLett.97.023602,PhysRevLett.114.095503}. The second mirror is a standard reflector that fully reflects light at frequency $\omega_2$. These optical fields interact with the atoms via a V-type coupling, which can be achieved by using circularly polarized light and placing a quarter-wave plate between the BEC and the mirrors. This setup ensures that there is no interference between the forward and backward propagating beams. Similar configurations have been used to engineer long-range two-body interactions~\cite{PhysRevLett.121.073604,PhysRevA.103.023308}, although previous studies only used a single monochromatic optical field. In the following, we demonstrate that the two-body effects can be entirely eliminated by using two pumping lasers with different frequencies. As a result, the dynamics of the condensate is governed by an effective three-body interaction. 

The two pumping beams first propagate through the quasi-2D BEC and then are reflected back from the mirrors, inducing feedback effects on the atoms. The forward (${\mathcal{E}}^{+}_{1,2}$) and backward (${\mathcal{E}}^{-}_{1,2}$) propagating fields couple the atomic ground state $\left|g\right\rangle $ and the two excited states $\left|\pm\right\rangle $ with the detunings $\Delta^+_{1,2}$ and $\Delta^-_{1,2}$, respectively. As discussed in Refs.~\cite{PhysRevLett.121.073604,PhysRevA.103.023308}, under the conditions of large detunings and a thin medium, the populations of excited-state atoms, as well as the diffraction of the optical fields within the condensate, are negligible. Consequently, the forward light fields acquire a phase shift depending on the local density of the BEC as below,
\begin{equation}
    {\mathcal{E}}^+_\sigma(\mathbf{r})= {\mathcal{E}}^{\rm in}_\sigma e^{-i\frac{k_{\sigma }\mu^2}{2\epsilon_0 \hbar \Delta^+_{\sigma }}|{\psi}(\mathbf{r})|^2 } \qquad (\sigma=1,2)
	\label{eq:1}
\end{equation}
with $\mathbf{r}=(x,y)$ representing the coordinates in the transverse plane perpendicular to the light propagating direction, ${\psi}(\mathbf{r})$ the wave function of ground-state atoms, ${\mathcal{E}}^{\rm in}_\sigma$ the constant amplitude of the incident plane-wave light, $k_\sigma$ the wave number of the corresponding optical field, and $\Delta^+_\sigma$ the detuning between the light and the atomic transition. Here, we have assumed equal dipole matrix elements $\mu$ for the two atomic transitions. The subsequent propagation of the light fields between the BEC and the mirrors is entirely governed by diffraction. Eventually, the amplitudes of the back-reflected beams incident on the BEC are given by
\begin{equation}
	{\mathcal{E}}^-_\sigma (\mathbf{r})=\frac{1}{2\pi}\int \tilde{\mathcal{E}}^{+}_\sigma (\mathbf{p}) e^{2d_\sigma \left( \sqrt{\mathbf{p}^2-k^2_\sigma}-ik_\sigma \right)}  e^{-i\mathbf{p}\cdot\mathbf{r}}  d^2\mathbf{p},
	\label{eq:2}
\end{equation}
where $\tilde{\mathcal{E}}^{+}_\sigma (\mathbf{p})$ is the Fourier transform of ${\mathcal{E}}^+_\sigma(\mathbf{r})$. The first exponential term in the above integral describes the diffraction of the light fields as they propagate between the BEC and the mirrors with a total distance of $2d_\sigma$.

In the case of large detunings, the light-induced dipole potential acting on the atoms reads
\begin{equation}
	{V}(\mathbf{r}) = \sum_{\sigma=1,2}\frac{\hbar}{4} \left( \frac{\left| \Omega^+_\sigma \right| ^2}{\Delta^+_\sigma } + \frac{\left| \Omega^-_\sigma \right| ^2}{\Delta^-_\sigma}\right) 
	\label{eq:3}
\end{equation}
with $\Omega^{\pm}_\sigma=\mu\mathcal{E}^{\pm}_\sigma /\hbar$ being the Rabi frequency of the light field. Substituting Eqs.~\eqref{eq:1} and~\eqref{eq:2} into Eq.~\eqref{eq:3}, the dipole potential can be rewritten as
\begin{equation}
	\begin{split}
	{V}&(\mathbf{r})
	=\sum_{\sigma=1,2}\frac{\hbar}{4} \bigg\{ \frac{\Omega^2_\sigma }{\Delta^+_\sigma} \\
	&+  \frac{\Omega^2_\sigma}{4\pi^2\Delta^-_\sigma} \left| \int e^{ -i \frac{k_{\sigma }\mu^2}{2\epsilon_0 \hbar \Delta^+_{\sigma }} |{\psi}(\mathbf{r}^{\prime})|^2 }  f_\sigma(\mathbf{r} - \mathbf{r}^{\prime})  d^2\mathbf{r}^{\prime}  \right|^2 \bigg\},
	\end{split}
	\label{eq:4}
\end{equation}
where $f_\sigma(\mathbf{r})={(2\pi)}^{-1}\int e^{2d_\sigma \left( \sqrt{\mathbf{p}^2-k^2_\sigma}-ik_\sigma\right)}  e^{-i\mathbf{p}\cdot\mathbf{r}} d^2\mathbf{p} $ and $\Omega_\sigma={\mu\mathcal{E}^{\rm in}_\sigma}/{\hbar}$. By expanding the exponent in the above equation and retaining terms up to second order in the density $|\psi|^2$, we ultimately obtain the following effective potential in the small phase limit 
$\frac{k_{\sigma }\mu^2}{2\epsilon_0 \hbar \Delta^+_{\sigma }} |{\psi}(\mathbf{r}^{\prime})|^2\ll 1$ 
(i.e., the thin medium condition),
\begin{equation}
{V}(\mathbf{r})={V}^{\rm II}(\mathbf{r})+{V}^{\rm III}(\mathbf{r}),
\label{eq:5}
\end{equation}
where
\begin{equation}
	{V}^{\rm II}(\mathbf{r})=-2\hbar\sum_{\sigma=1,2}\alpha_\sigma \beta_\sigma  \frac{k_\sigma}{d_\sigma} \int \left| {\psi}(\mathbf{r}^{\prime}) \right| ^2 \cos\left(\frac{k_\sigma (\mathbf{r} - \mathbf{r}^{\prime})^2}{4d_\sigma}\right) d^2\mathbf{r}^{\prime}
	\label{eq:6}
\end{equation}
and
\begin{equation}
	\begin{split}
	{V}^{\rm III}&(\mathbf{r})=\hbar\sum_{\sigma=1,2} \alpha_\sigma \beta^2_\sigma \bigg\{  \frac{1}{4}  \left| \frac{k_\sigma}{d_\sigma} \int \left| {\psi}(\mathbf{r}^{\prime}) \right| ^2 e^{i\frac{k_\sigma (\mathbf{r} - \mathbf{r}^{\prime})^2}{4d_\sigma}} d^2\mathbf{r}^{\prime} \right| ^2\\
	&- \pi \frac{ k_\sigma}{d_\sigma} \int \left| {\psi}(\mathbf{r}^{\prime}) \right| ^4 \sin\left(\frac{k_\sigma (\mathbf{r} - \mathbf{r}^{\prime})^2}{4d_\sigma}\right) d^2\mathbf{r}^{\prime} \bigg\}
	\end{split}
	\label{eq:7}
\end{equation}
correspond to the light-induced two-body and three-body interactions, respectively. Here, we have defined $\alpha_\sigma \equiv \Omega^2_\sigma/16\pi^2\Delta^-_\sigma$, 
$\beta_\sigma \equiv k_{\sigma }\mu^2/2\epsilon_0 \hbar \Delta^+_{\sigma }$, 
and omitted the constant term. To obtain the analytical formulation of these effective interactions, we have assumed that $d_\sigma \gg k^{-1}_\sigma$, and thus $ f_\sigma (\mathbf{r}) \approx -i\frac{k_\sigma}{2d_\sigma}  e^{i\frac{k_\sigma}{4d_\sigma} r^2 }$~\cite{PhysRevLett.121.073604,PhysRevA.103.023308}.

Now that the optical fields have been eliminated in the expression of the dipole potential, resulting in effective multi-body interactions between the atoms, it becomes possible to describe the dynamics of the ground-state atoms using the following closed Gross-Pitaevskii equation (GPE),
\begin{equation}
	i\hbar \frac{\partial {\psi}(\mathbf{r})}{\partial t } =
	-\frac{\hbar^2\nabla_{\bot}^2  }{2m} {\psi}(\mathbf{r}) +g_{\rm 2D} |{\psi}(\mathbf{r})|^2 {\psi}(\mathbf{r}) + {V}(\mathbf{r})  {\psi}(\mathbf{r}),
	\label{eq:8}
\end{equation}
where $\nabla_{\bot}^2=\frac{\partial^2}{\partial x^2} +\frac{\partial^2}{\partial y^2}$, $m$ is the atomic mass, and $g_{\rm 2D}=\frac{4\pi\hbar^2 a_s}{m}\int |\varphi(z)|^4 dz$ represents the quasi-2D interaction strength with $a_s$ the $s$-wave scattering length and $\varphi(z)$ the wave function of the quasi-2D BEC along the strong confinement direction~\cite{PhysRevLett.84.2551}. This GPE provides a flexible framework for exploring many-body physics related to two-body interactions, three-body interactions, and their competition. Not only can the contact two-body interaction be finely tuned across a wide range via Feshbach resonance, but also the light-induced interactions are highly controllable. The relevant parameters, $\alpha_\sigma$ and $\beta_\sigma$, can be independently adjusted by varying the Rabi frequencies $\Omega_\sigma$ or the detunings $\Delta^+_\sigma$. Most particularly, in contrast to previous schemes where synthetic three-body interactions are typically accompanied by stronger two-body effects, the light-induced two-body interaction $V^{\rm II}$ in our proposal can be completely switched off, leaving only the three-body interaction $V^{\rm III}$.

The effects of the light-induced two-body interaction $V^{\rm II}$ have been explored in our previous studies~\cite{PhysRevLett.121.073604,PhysRevA.103.023308}, here we will focus on the three-body interaction $V^{\rm III}$. We will do this by eliminating the two-body effect, which can be achieved by a proper choice of parameters, specifically,
\begin{equation}
	\frac{k_1}{d_1}=\frac{k_2}{d_2}, \qquad \alpha_1=\alpha_2=\alpha, \qquad \beta_1=-\beta_2=\beta.
	\label{eq:9}
\end{equation}
This will make $\alpha_1\beta_1+\alpha_2\beta_2=0$, and hence the light-induced two-body interaction vanishes [cf. Eq.~\eqref{eq:6}], while the three-body interaction $\alpha_1\beta_1^2+\alpha_2\beta_2^2=2\alpha\beta^2$ remains finite. In this case, the GPE in Eq.~\eqref{eq:8} reduces to 
\begin{equation}
	i\frac{\partial {\psi}(\mathbf{r})}{\partial t } =
	-\frac{\nabla_{\bot}^2 }{2} {\psi}(\mathbf{r}) +g N |{\psi}(\mathbf{r})|^2 {\psi}(\mathbf{r}) + \tilde{V}^{\rm III}(\mathbf{r})  {\psi}(\mathbf{r}),
	\label{eq:10}
\end{equation}
containing only the contact two-body interaction and the effective three-body interaction,
\begin{equation}
\begin{split}
	\tilde{V}^{\rm III}(\mathbf{r}) =&\frac{\mathcal{C}_3 N^2 }{4} \bigg\{  \left| \int \left| {\psi}(\mathbf{r}^{\prime}) \right| ^2 e^{i \frac{(\mathbf{r}-\mathbf{r}')^2}{4}} d^2\mathbf{r}^{\prime} \right| ^2\\
	 &-4\pi  \int \left| {\psi}(\mathbf{r}^{\prime}) \right| ^4 \sin{\frac{(\mathbf{r}-\mathbf{r}')^2}{4} } d^2\mathbf{r}^{\prime} \bigg\},
\end{split}
	\label{eq:11}
\end{equation}
with $\mathcal{C}_3= 2\alpha\beta^2 \frac{mk}{\hbar d}$ describing the three-body interaction strength, and $g=g_{\rm 2D}\frac{m}{\hbar^2}$. Here, space and time have been rescaled by $\sqrt{\frac{d_1}{k_1}}$ and $\frac{md_1}{\hbar k_1}$, respectively, giving the above dimensionless Eqs.~\eqref{eq:10} and~\eqref{eq:11}. For a typical setting of rubidium BEC with $\lambda_1=780{\rm nm}$, $d_1=10\lambda_1$, these two scales are given by $\sqrt{\frac{d_1}{k_1}} \sim 1\mu$m and $\frac{md_1}{\hbar k_1}\sim 1.3 {\rm ms}$~\cite{PhysRevLett.121.073604}. Moreover, $N$ represents the total number of particles in the BEC with the normalized wave function $\int |\psi (\mathbf{r})|^2 d^2\mathbf{r}=1$. 

To gain an intuitive understanding of the properties of this three-body interaction, we reformulate Eq.~\eqref{eq:11} into the following equivalent expression,
\begin{equation}
	\begin{split}
		\tilde{V}^{\rm III}&(\mathbf{r}) =\frac{\mathcal{C}_3 N^2}{4}  \bigg\{  \int \left| {\psi}(\mathbf{r}^{\prime}) \right| ^2 U_c(\mathbf{r},\mathbf{r}',\mathbf{r}'')\left| {\psi}(\mathbf{r}^{\prime\prime}) \right| ^2 d^2\mathbf{r}^{\prime} d^2\mathbf{r}^{\prime \prime}\\
		& -4\pi  \int \left| {\psi}(\mathbf{r}^{\prime}) \right| ^2  U_s(\mathbf{r},\mathbf{r}',\mathbf{r}'') \left| {\psi}(\mathbf{r}^{\prime\prime}) \right| ^2 d^2\mathbf{r}^{\prime} d^2\mathbf{r}^{\prime\prime} \bigg\},
	\end{split}
	\label{eq:12}
\end{equation}
where we have defined $U_c(\mathbf{r},\mathbf{r}',\mathbf{r}'')\equiv \cos{ \frac{(\mathbf{r}-\mathbf{r}')^2-(\mathbf{r}-\mathbf{r}'')^2}{4}}$, and $U_s(\mathbf{r},\mathbf{r}',\mathbf{r}'')\equiv \sin{\frac{(\mathbf{r}-\mathbf{r}')^2}{4} } \delta(\mathbf{r}'-\mathbf{r}'')$. Eq.~\eqref{eq:12} describes the interaction potential at position $\mathbf{r}$ generated by two particles at positions $\mathbf{r}'$ and $\mathbf{r}''$, respectively, implying that a finite force between them requires the presence of at least three particles.
It is important to note that, on the one hand, this three-body interaction $\tilde{V}^{\rm III}$ possesses a typical long-range character, similar to the two-body interaction $V^{\rm II}$; on the other hand, $\tilde{V}^{\rm III}$ is asymmetric when swapping $\mathbf{r}$ with either $\mathbf{r}'$ or $\mathbf{r}''$, unlike $V^{\rm II}$, which maintains pairwise symmetry. Furthermore, based on the expressions of $U_c(\mathbf{r},\mathbf{r}',\mathbf{r}'')$ and $U_s(\mathbf{r},\mathbf{r}',\mathbf{r}'')$, it is evident that the former can produce a finite contribution to the three-body potential $\tilde{V}^{\rm III}$ in the scenario of three separated particles, while the latter has zero contribution in this configuration. This is due to the hybrid nature of $U_s(\mathbf{r},\mathbf{r}',\mathbf{r}'')$, which integrates both long-range and contact properties. To obtain a finite contribution from $U_s$, two of the three particles must occupy the same position (that is, $\mathbf{r}'=\mathbf{r}''$). 

If we just consider three particles under the influence of the three-body potential $U_c$ and $U_s$, the only stable configuration is that the particles occupy the three vortices of an equilateral triangle, for which case, the force on each particle is zero. This turns out to have important implications for pattern formations in a BEC (cf. Sec.~\ref{sec:3}). Surprisingly, the total force becomes finite in asymmetric configurations, demonstrating the remarkable nonreciprocal characteristic of the three-body interaction. These long-range and nonreciprocal properties distinguish it significantly from synthetic three-body interactions proposed in lattice models~\cite{Johnson_2009,PhysRevA.82.043629,PhysRevA.78.043603} or binary quantum gases~\cite{PhysRevLett.112.103201,PhysRevLett.128.083401}. In what follows, we discuss the stationary and dynamical behaviors of the condensate governed by this induced three-body interaction.

\section{Stable stationary states in the condensate}
\label{sec:3}

\begin{figure}[!b]
	\centering
	\includegraphics[width=\columnwidth]{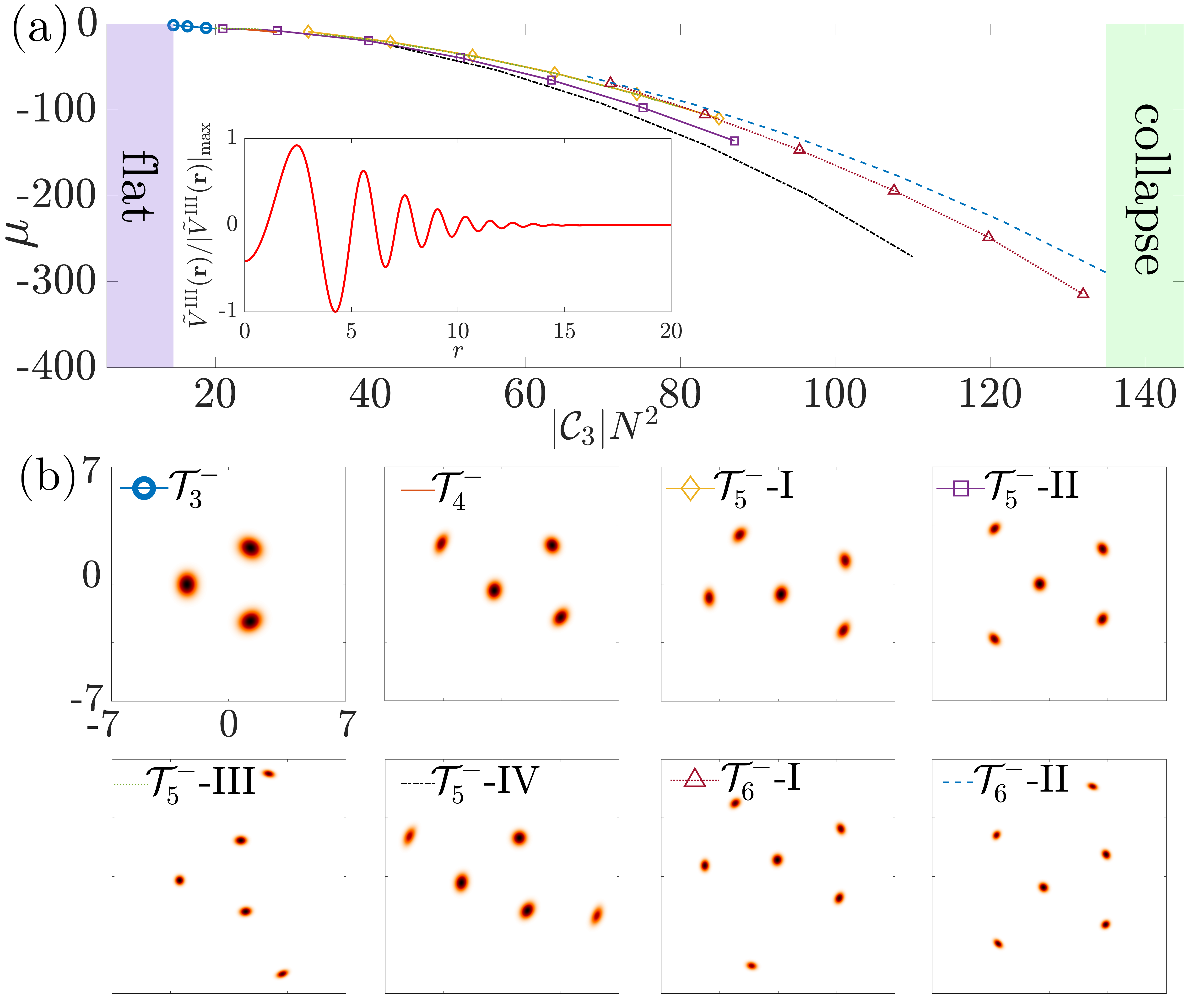}
	\caption{(a) The chemical potentials of the stable stationary states vary with the three-body interaction strength $|\mathcal{C}_3|N^2$ in the situation of $\mathcal{C}_3<0$ and $g=0$. The density profiles of the corresponding states are exhibited in (b). The inset in (a) presents the landscape of the three-body potential generated by a Gaussian wave function $e^{-(x^2+y^2)/w^2}$ with $w=0.9$.}
	\label{fig2}
\end{figure}

Let us start analyzing the light-induced three-body interaction by exploring stationary-state properties of the BEC for a fixed particle number $N$. To obtain stable stationary states, we use the imaginary-time evolution approach. Specifically, we propagate the GPE~\eqref{eq:10} in imaginary time (i.e., replacing $t$ with $-it$) starting from a random state and renormalize the wave function at each step. This process eventually relaxes the condensate to a stable state. Since this method is sensitive to the initial condition, the final converged state could be either the ground state with the lowest energy or a metastable state with locally minimum energy. However, due to the nonreciprocal nature of the effective three-body interaction, i.e., non-Hamiltonian~\cite{PhysRevX.5.011035}, there is no well-defined energy in this system. In other words, it is impossible to identify the ground state by comparing the energies of the converged states, as is typically done. However, it is still possible to explore the stationary properties of the condensate using the imaginary-time evolution algorithm.

According to Eq.~\eqref{eq:11}, we observe that the three-body interaction coefficient $\mathcal{C}_3$ has the same sign as $\alpha$, and more specifically the sign of detuning $\Delta^-_\sigma$. Therefore, it allows us to flip the sign of the three-body interaction by adjusting the detuning between the light and the atoms. Let us first consider the case where $\mathcal{C}_3<0$ with vanishing contact interaction (i.e., $g=0$). For a relatively weak three-body interaction, the distribution of the BEC is primarily dominated by the kinetic energy term, which causes the atoms to spread out, forming an extended flat state. However, as the three-body interaction increases beyond the threshold $|\mathcal{C}_3| N^2\sim 15$, as shown in Fig.~\ref{fig2}, the three-body interaction can stabilize the condensate into various localized droplet cluster states. Despite the noticeable differences between the profiles of these droplets, they all share a common fundamental structure: an equilateral triangle composed of three droplets. This three-droplet cluster [see the subplot in Fig.~\ref{fig2}(b)] represents the smallest-size stationary state supported by this three-body interacting system. This is in sharp contrast to the two-body interaction $V^{\rm II}$, which can stabilize the condensate into a single droplet~\cite{PhysRevLett.121.073604}. To understand the origin of these cluster states, we plot the three-body potential of a single Gaussian state [see the inset in Fig.~\ref{fig2}(a). This potential features an oscillatory landscape, reaching its minima at $r=0$ and $\delta^-_k\approx\sqrt{2\pi(4k-1)}$ ($k=1,2,3,\cdots$). Surprisingly, although $\tilde{V}^{\rm III}$ presents an attractive potential when particle separation approaches zero [see the potential minimum at the zero point], neither a single-droplet nor a two-droplet state can be stabilized. This is because the second minimum at $\delta^-_1\approx 4.3$ is lower than the minimum value at $r=0$. As a result, the condensate tends to divide and eventually self-organize into these droplet clusters. Moreover, one can notice that the side length of the basic triangle is approximately equal to $\delta^-_1$. If each droplet is considered as a super-particle, this minimum-size three-droplet state reflects a key characteristic of the three-body interaction, i.e., it requires at least three particles to maintain a finite interaction, otherwise, the three-body potential vanishes.

As the three-body interaction increases, the three-droplet state loses its stability, and other stable cluster sates composed of more droplets emerge. To understand this proliferation behavior, we examine the three-body potential generated by the basic three-droplet cluster $\mathcal{T}^-_3$. There are a number of local potential minima surrounding the central three droplets. These local minima deepen as $|\mathcal{C}_3|N^2$ increases, allowing for the addition of another droplet and forming multi-droplet clusters with various structures, as illustrated in Fig.~\ref{fig2}(b). However, these self-bound cluster states eventually terminate at sufficiently strong three-body interactions, where the condensate tends to collapse. We would like to point out that this collapse differs from the usual collapse driven by attractive contact interactions~\cite{Wieman2001Nature} or anisotropic dipole-dipole interactions~\cite{Pfau2009review}, where the atoms contract to a singular point in space. Due to the non-monotonic oscillating nature of the three-body interaction, it favors the formation of multiple, separated condensate pieces with divergent densities rather than a singular point [cf. Fig.~\ref{fig4}(e4)]. This \textit{diffusive collapse} is a unique behavior caused by this unconventional long-range three-body interaction and, to our knowledge, has not been reported in other systems.

The stability of these droplet cluster states is also sensitive to contact interaction. We have also examined the corresponding stable regime of each droplet cluster state in the presence of contact repulsion and found that both the lower and upper critical points of the stable region shift towards larger $|\mathcal{C}_3| N^2$ with the introduction of a finite $gN$. However, as the contact interaction increases, the stable regions shrink and eventually vanish.

\begin{figure}[!b]
\centering
\includegraphics[width=\columnwidth]{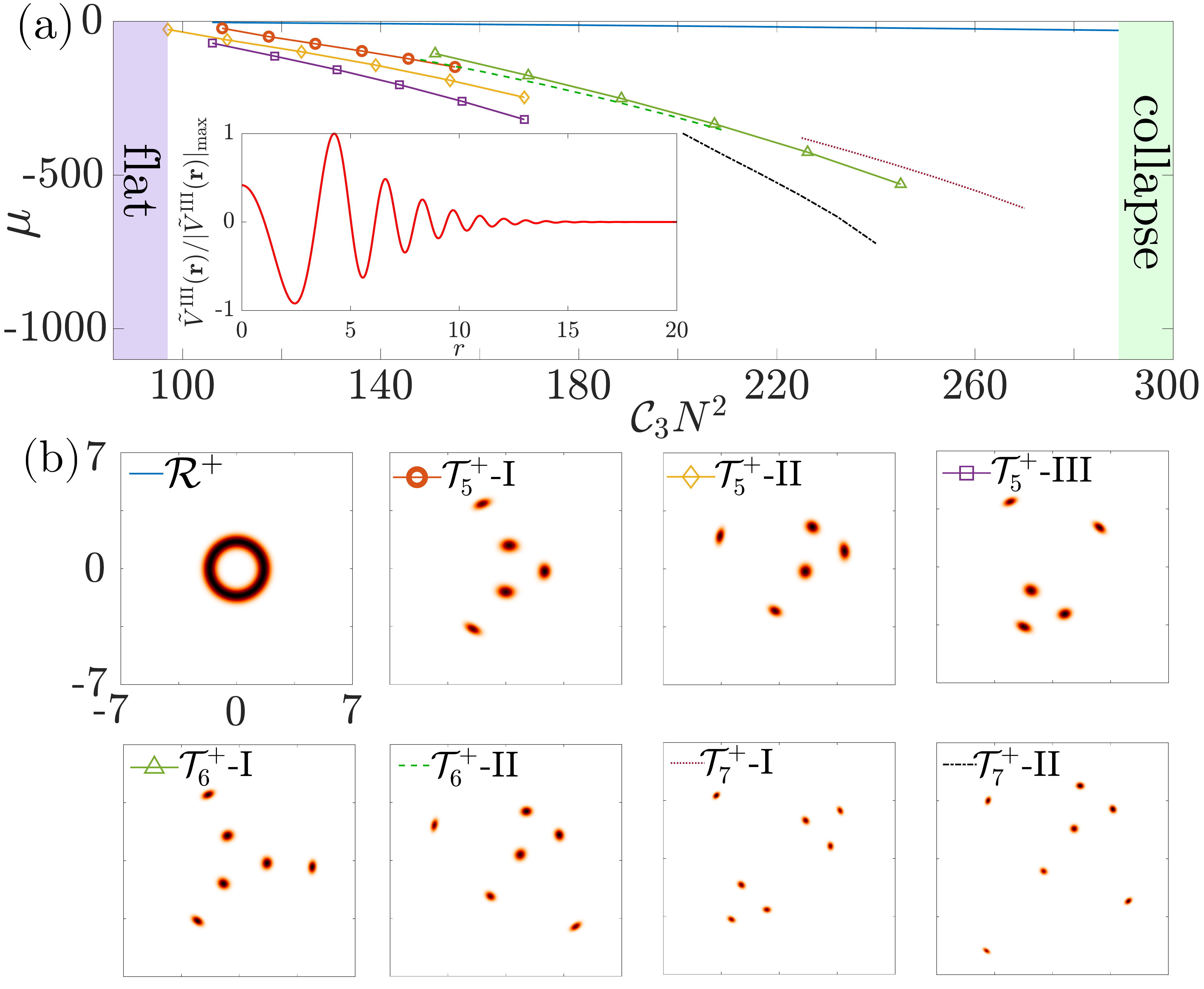}
\caption{(a) The chemical potentials of the stable stationary states vary with the three-body interaction strength $\mathcal{C}_3N^2$ in the situation of $\mathcal{C}_3>0$ and $g=0$. The density profiles of the corresponding states are exhibited in (b). The inset in (a) presents the landscape of the three-body potential generated by a Gaussian wave function $e^{-(x^2+y^2)/w^2}$ with $w=0.9$.}
\label{fig3}
\end{figure}

Next, we turn to consider the positive case of $\mathcal{C}_3>0$ at $g=0$. In contrast to the negative case discussed above, as shown by the inset in Fig.~\ref{fig2}(a), the three-body interaction now exhibits short-range repulsion and reaches its minimum potential at $\delta^+_k=\sqrt{2\pi(4k-3)}$ ($k=1,2,3,\cdots$) [see the inset in Fig.~\ref{fig3}(a)]. Due to the repulsive interaction at zero separation, the condensate tends to spread rather than contract. Meanwhile, the long-range attraction prevents the condensate from spreading. Consequently, the BEC can also stabilize into a series of stationary states, including an emergent self-bound ring state and various droplet clusters, as shown in Fig.~\ref{fig3}(b). Similarly to the negative case discussed before, the fundamental structure of these cluster states remains an equilateral triangle, but with a different side length. However, the three-droplet cluster state no longer exists in this case. Instead, the smallest stable cluster state consists of five droplets [see Fig.~\ref{fig3}(b)]. By analyzing the density distribution of the $\mathcal{T}^+_6{\rm -I}$ state, one can observe that the separation between the three central droplets is equal to $d\approx 3$. It is quite close to $\delta^+_1\approx 2.5$ where the three-body interaction attains its first minimum, as illustrated by the inset in Fig.~\ref{fig3}(a). The slight mismatch between these two length scales is due to the competition between the multi-droplets.

Remarkably, the condensate can form a stable ring state $\mathcal{R}^+$ over a much broader region than the cluster states. It should be noted that the radius of this ring state is $r=1.7\approx d/\sqrt{3}$, indicating that the ring corresponds to the circumcircle of the equilateral triangle with the side length of $d\approx 3$ (i.e., the central three droplets of the $\mathcal{T}^+_6{\rm -I}$ state). To obtain a ring state in BECs, it is usually based on a toroidal potential~\cite{PhysRevLett.99.260401,PhysRevA.79.043620}. In contrast, our setup does not use any external trap in the transverse direction perpendicular to the light propagation. This means that the ring state, as well as the droplet clusters, are all self-bound states solely led by the effective three-body interaction, offering a new opportunity to explore interesting physics, for example, persistent flow~\cite{PhysRevLett.99.260401,PhysRevA.95.041604}, vortex dynamics~\cite{PhysRevA.79.043620}, quantized superfluid~\cite{Eckel2014Nature}, and symmetry breaking~\cite{PhysRevA.95.033620}, based on self-organized annular BECs. 

When comparing the two positive and negative cases of $\mathcal{C}_3$, another dramatic difference is the stability of the self-bound sates in the presence of finite contact repulsion. The stable region of each state undergoes a noticeable expansion with increasing contact interaction, which contrasts sharply with the behavior observed in the positive case (not shown).

\section{Dynamics of the condensate}
\label{sec:4}

In this section, we explore the dynamics of the BEC driven by the effective three-body interaction in the absence of contact interaction (i.e., $gN=0$). Here, we consider the real-time evolution of the condensate starting from two different types of initial state: (1) symmetric states, where the spatial distribution symmetries are maintained, and (2) asymmetric states, where the symmetry is broken. In Sec.~\ref{sec:4a}, we first investigate the dynamics starting from symmetric states and show that various intermediate states, featuring similar profiles of the aforementioned stable stationary states, emerge during evolution. Subsequently, we turn to the situation of asymmetric initial states in Sec.~\ref{sec:4b}, and unveil an intriguing self-acceleration phenomenon that manifestly violates Newton's law of motion.

\subsection{Dynamical self-organization from symmetric initial states}
\label{sec:4a}

\begin{figure}[!t]
	\centering
	\includegraphics[width=\columnwidth]{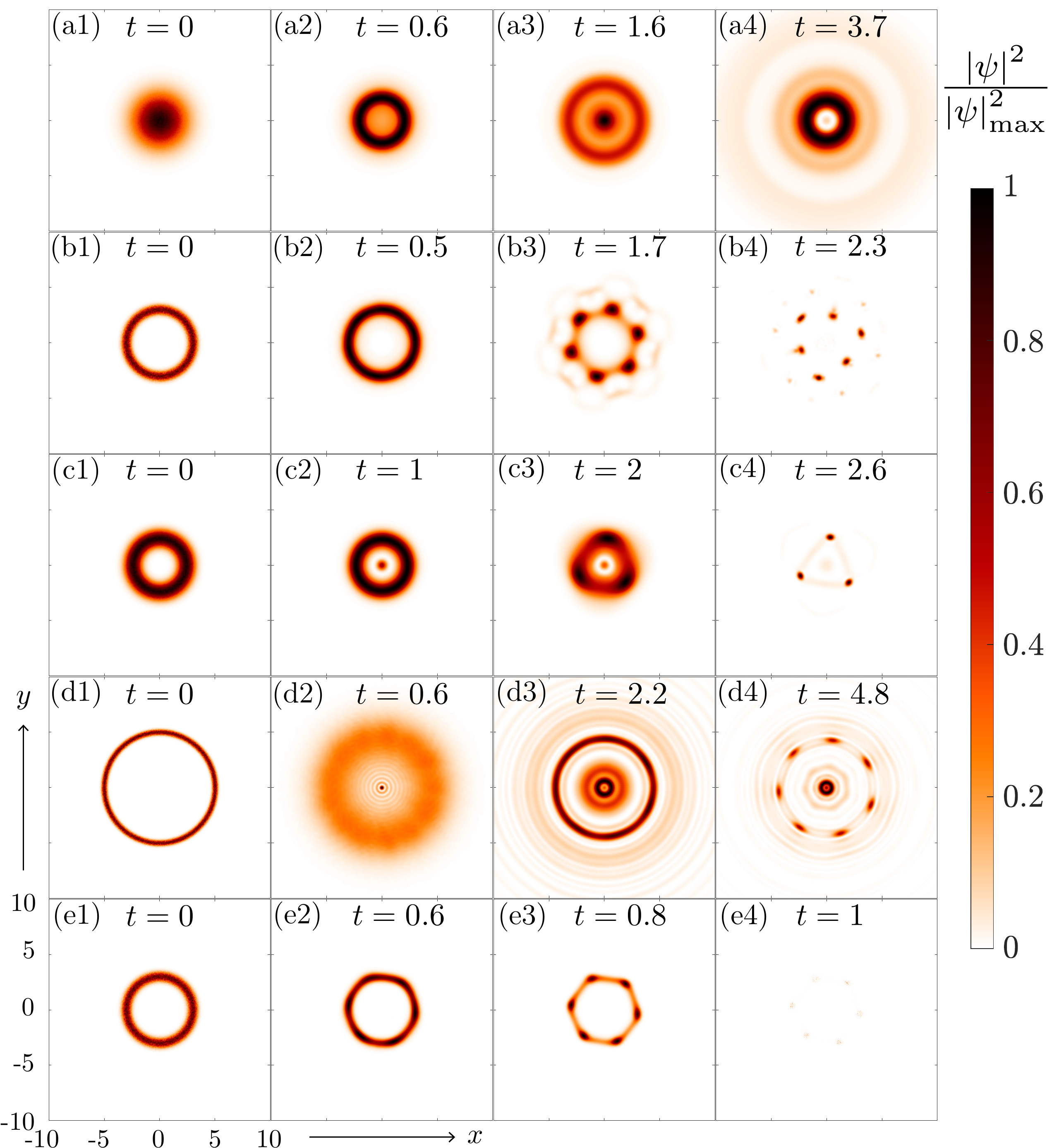}
	\caption{Snapshots of the condensate's density profiles during the real-time evolution starting from symmetric initial states, i.e., a Gaussian state (a) and ring states (b$\sim$d) with different radii, respectively. Here, the contact interaction has been turned off (i.e., $gN=0$), and the corresponding three-body interaction strengths $\mathcal{C}_3N^2$ are (a) $80$, (b) $280$, (c) $=-120$, (d) $-360$, and (e) $500$. }
	\label{fig4}
\end{figure}

One standard method for experimentally observing the ring state and droplet clusters discussed in Sec.~\ref{sec:3} is examining the dynamics of the system starting from certain initial states that are easy to prepare. To simulate this process, we assume either a Gaussian or a ring-shaped initial wave function and propagate the GPE, i.e., Eq.~\eqref{eq:10}, in real time. Let us first consider the situations of positive three-body interactions. Figs.~\ref{fig4}(a1$\sim$a4) provide an example of the real-time evolution starting from a Gaussian state at $\mathcal{C}_3 N^2=80$. Notably, the BEC gradually organizes into an annular profile as time increases, reminiscent of the ring state discussed earlier. We have also explored the dynamics starting from ring-shaped states as shown in Fig.~\ref{fig4}(b). For the case of an initial radius $r=3$ as exhibited in Figs.~\ref{fig4}(b1$\sim$b4), where the three-body interaction is fixed at $\mathcal{C}_3 N^2=280$, the condensate exhibits a gradual fission along the angular direction and eventually transforms into a cluster state composed of six droplets around $t\sim 2.3$. In contrast, the condensate rapidly collapses into isolated singular points at sufficiently strong three-body interaction as shown in Fig.~\ref{fig4}(e1$\sim$e4) for $\mathcal{C}_3 N^2=500$, where the initial state is the same as in Fig.~\ref{fig4}(b). 

Figs.~\ref{fig4}(c,d) present the dynamics in the case of negative three-body interactions. For the initial state with a small radius of $r=2.4$ at $\mathcal{C}_3 N^2=-120$ [see Figs.~\ref{fig4}(c1$\sim$c4)], the condensate dynamically self-organizes into a three-droplet cluster state, the basic profile of the stationary states. By increasing the initial radius to $r=5$ and tuning the three-body interaction to $\mathcal{C}_3 N^2=-360$, the condensate undergoes a typical self-interference process~\cite{Lannert2007PRA,Tononi2020PRL,Wang2022PRL} during the early stages of evolution, resulting in a central density peak appearing at $t\sim 0.6$. After that, the condensate continues to evolve, and a peculiar state with seven-fold rotational symmetry emerges as shown in Figs.~\ref{fig4}(d1$\sim$d4).

\subsection{Self-acceleration from asymmetric initial states}
\label{sec:4b}

\begin{figure}[!b]
\centering
\includegraphics[width=\columnwidth]{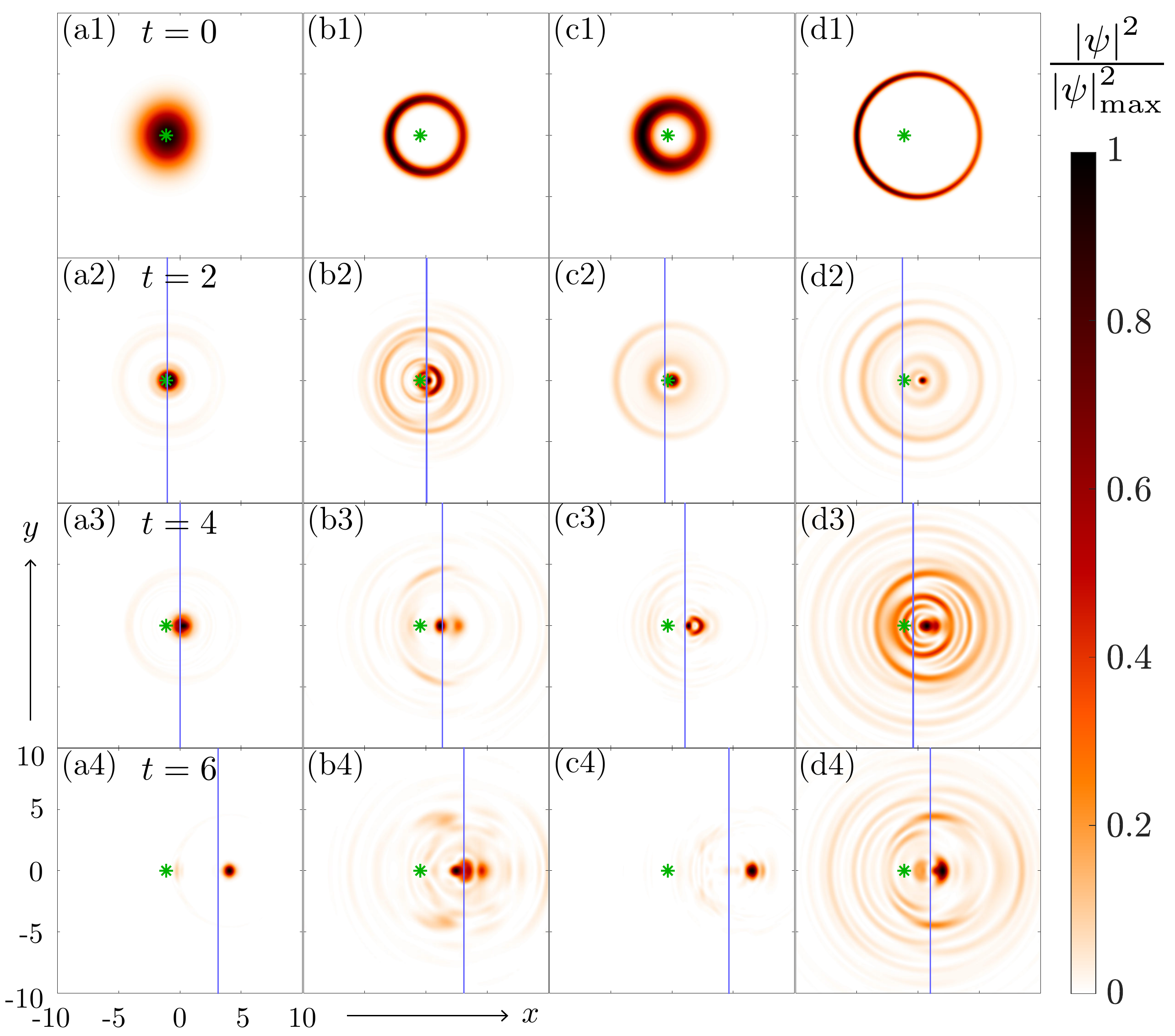}
\caption{Snapshots of the condensate's density profiles during the real-time evolution starting from asymmetric initial states. Here, we add a gradient along $x$-direction to the same symmetric Gaussian and ring states used in Fig.~\ref{fig4} by multiplying symmetric states with the tilt factor $10-(x-r)$ and then renormalizing them. The green asterisk and blue vertical line mark the COM positions of the initial state and the instantaneous state, respectively. The simulations in (a$\sim$d) are conducted using the effective three-body interactions Eq.~\eqref{eq:7}. The respective three-body interaction strengths $\mathcal{C}_3N^2$ are (a) $-40$, (b) $-160$, (c) $=-80$, and (d) $-360$.}
\label{fig5}
\end{figure}

In addition to the various self-organized profiles that arise in the BEC's evolution from a symmetric initial state, it is also worth mentioning another characteristic of the dynamics, i.e., the center of mass (COM) of the entire condensate remains stationary throughout the whole evolution process. According to Newton's law of motion, this is expected since the initial velocity is zero and no external force is acting on the BEC. Although we shine light on the atoms, the optical fields are ultimately eliminated under the condition of large detunings, yielding an effective many-body model [see the GPE in Eq.~\eqref{eq:10}]. Clearly, Eq.~\eqref{eq:10} describes a closed system without any external potential, involving only the kinetic energy and internal interactions between atoms. Therefore, it seems reasonable to expect an immobile COM in the case of a vanishing velocity at the beginning, regardless of the profiles of the initial states. 

However, when the evolution begins from an asymmetric initial state, the COM of the condensate gradually acquires a finite velocity, accompanying the dynamical self-organization. For example, as shown in Fig.~\ref{fig5}, we introduce an overall gradient along $x$-direction to the same initial states used in Fig.~\ref{fig4}, breaking the spatial distribution symmetry. In sharp contrast to the symmetric case (cf. Fig.~\ref{fig4}), one can note that the COM of the condensate (denoted by the blue vertical lines in Fig.~\ref{fig5}) moves towards the right. Such a counterintuitive behavior arises from the intriguing non-reciprocity of the light-induced three-body interaction. Although Newton's law of motion is generally obeyed in systems with reciprocal interactions (i.e., action $=$ reaction), it can, in principle, be violated when such reciprocity is broken. For the effective three-body interaction considered here, for a simple system consisting of only three particles, the force acting on each particle $\mathbf{F}_j$ is highly dependent on the spatial configuration. When the three particles are placed, e.g., at the vertices of an equilateral triangle, the total force vanishes. That is, $\mathbf{F}_1+\mathbf{F}_2+\mathbf{F}_3=0$. In contrast, when the positions of the three particles lose regular symmetry, the total force becomes finite, driving the system to move. Several schemes have been proposed to effectively generate non-reciprocal two-body interactions in classical systems including, e.g., biological systems~\cite{Kryuchkov2018,PhysRevX.12.010501} and soft matters~\cite{PhysRevLett.128.048002}. Our proposal presents a novel pathway to engineering nonreciprocal three-body interactions in quantum gases. Here, we have focused mainly on the negative case (i.e., $\mathcal{C}_3N^2<0$). Although self-acceleration behavior exists in the positive case as well, the condensate would tend to quickly spread in the early stage of the evolution, effectively reducing the atomic interactions and thereby leading to a smaller displacement of the COM. 

To obtain the effective three-body interaction, we have neglected all higher-order terms as demonstrated in Sec.~\ref{sec:2}. In order to verify that the self-acceleration of the condensate is not an artifact caused by this approximation, we also examined the dynamics using the full potential in Eq.~\eqref{eq:4}, which includes all contributions from both three-body and higher-order interactions. By starting the evolution of the BEC from the same initial state as in Fig.~\ref{fig5}(c1), but with the three-body potential $\tilde{V}^{\rm III}$ in Eq.~\eqref{eq:10} replaced by the full potential [see Eq.~\eqref{eq:4}, noting that the light-induced two-body effect remains vanishing under the condition in Eq.~\eqref{eq:9}], we confirmed that the self-acceleration of the condensate persists (not shown), indicating that this significant phenomenon can indeed be observed in realistic systems.

\begin{figure}[!b]
	\centering
	\includegraphics[width=\columnwidth]{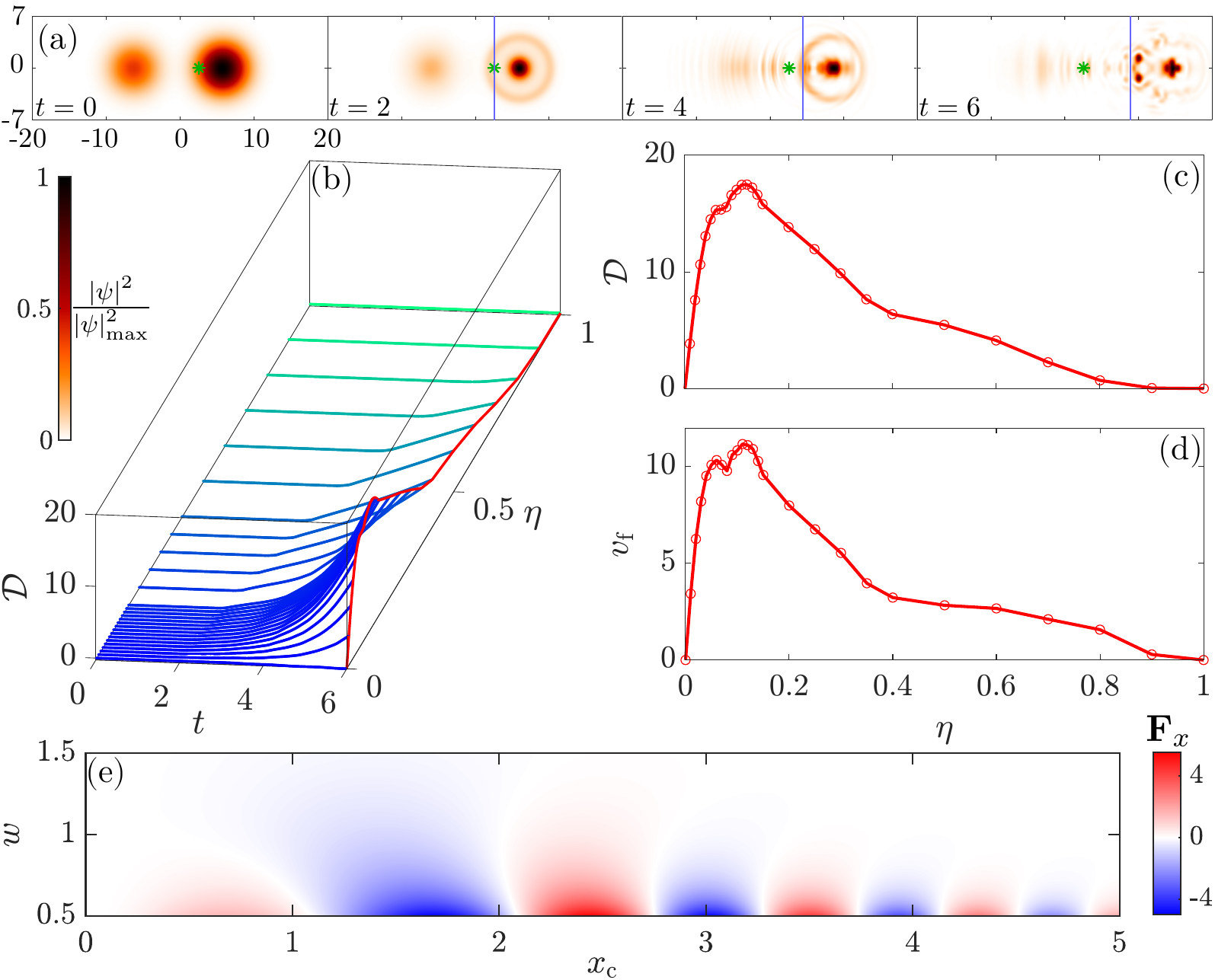}
	\caption{(a) Snapshots of the BEC's density profiles during the real-time evolution starting from the imbalanced two-Gaussian state in Eq.~\eqref{eq:13} at $\eta=0.4$ and $\mathcal{C}_3 N^2=-80$. Here, the green asterisk and blue vertical line mark the COM positions of the initial state and the instantaneous states, respectively. (b) The displacement $\mathcal{D}$ between the initial COM and instantaneous COM varies with time and $\eta$. (c) and (d) show the final displacement and velocity, respectively, at $t=6$ as a function of $\eta$. (e) exhibits the total force $F_x$ imposed on an asymmetric density profile consisting of two Gaussian functions, i.e., Eq.~\eqref{eq:13}, varying with the separation $x_c$ and width $w$ when $\eta=1/3$.}
	\label{fig6}
\end{figure}

From Fig.~\ref{fig5}, one can also note that the migration velocity of the COM significantly depends on the structure and size of the initial state. To gain a deeper insight into the relation between the self-acceleration and the initial conditions, we have explored the dynamics using other initial states, such as asymmetric super-Gaussian and imbalanced two-Gaussian states, in addition to the asymmetric Gaussian and ring-shaped states. In all cases, self-acceleration is present. This indicates that the self-acceleration behavior is independent of the specific initial state and is solely driven by the asymmetry of the initial condition and the non-reciprocity of the three-body interaction. In the following, we take the imbalanced two-Gaussian state as an example to analyze the relation between the self-acceleration and the asymmetry degree of the initial state. The density profile of such a state is given by
\begin{equation}
	\rho(\mathbf{r})=\frac{1}{(1+\eta)\pi w^2}\left[\eta e^{-\frac{(x+x_c)^2+y^2}{w^2}}+ e^{-\frac{(x-x_c)^2+y^2}{w^2}}\right],
	\label{eq:13}
\end{equation}
where $w$ is the width of the Gaussian, and the two Gaussian states are positioned at $(-x_c, 0)$ and $(x_c, 0)$, respectively, with a separation $2x_c$ along the $x$-direction. Moreover, $\eta$ represents the relative amplitude of the Gaussian on the left side compared to the Gaussian on the right side, allowing us to control the degree of asymmetry. As $\eta$ varies from 0 to 1, the degree of asymmetry first increases from a symmetric single Gaussian state and then decreases until it reaches a symmetric two-Gaussian state. Fig.~\ref{fig6}(e) shows the total force $\mathbf{F}_x=\int [-\nabla_x \tilde{V}^{\rm III}(\rho(\mathbf{r}))]\rho(\mathbf{r}) d^2\mathbf{r}$ acting on the COM of such a state when $\eta=1/3$. Notably, the total force vanishes at $x_c=0$, where the system is reduced to a symmetric state. However, the force attains a finite value and oscillates with the separation $x_c$. This non-vanishing force drives the movement of the condensate even in the absence of an external potential. Additionally, the total force gradually diminishes as $w$ or $x_c$ increases, since the two separated Gaussian states eventually merge into a single symmetric Gaussian state as $w$ diverges, or lose mutual interaction when their separation becomes too large.

By tuning $\eta$, we investigate the displacement $\mathcal{D}$ between the initial position of the COM and its instantaneous position. For example, we show the dynamics of such a two-Gaussian state at $\eta=0.4$ in Fig.~\ref{fig6}(a). Fig.~\ref{fig6}(b) shows the displacement $\mathcal{D}$ varying over time for different $\eta$. It can be observed that $\mathcal{D}$ maintains zero at $\eta=0$ and $1$, indicating that the COM of the condensate does not move at all. This is consistent with the stationary behavior of the COM in the case of a symmetric initial state as discussed in Sec.~\ref{sec:4a}. In contrast, the displacement $\mathcal{D}$ gradually increases with time when $0<\eta<1$. In Fig.~\ref{fig6}(c), we present the final displacement, within a fixed evolution time of $t=6$, as a function of $\eta$ [see the red line in Fig.~\ref{fig6}(b)]. Similarly to the degree of asymmetry mentioned above, the displacement also increases first with $\eta$. After $\mathcal{D}$ reaches its maximum value at $\eta \sim 0.12$, the displacement begins to decrease as $\eta$ increases. Furthermore, the final velocity $v_f$ of the COM presents behaviors similar to those shown in Fig.~\ref{fig6}(d). This demonstrates that self-acceleration is highly influenced by the degree of asymmetry. Such an asymmetry-dependent self-acceleration behavior is a unique characteristic of this nonreciprocal three-body interaction.

\begin{figure}[!t]
	\centering
	\includegraphics[width=\columnwidth]{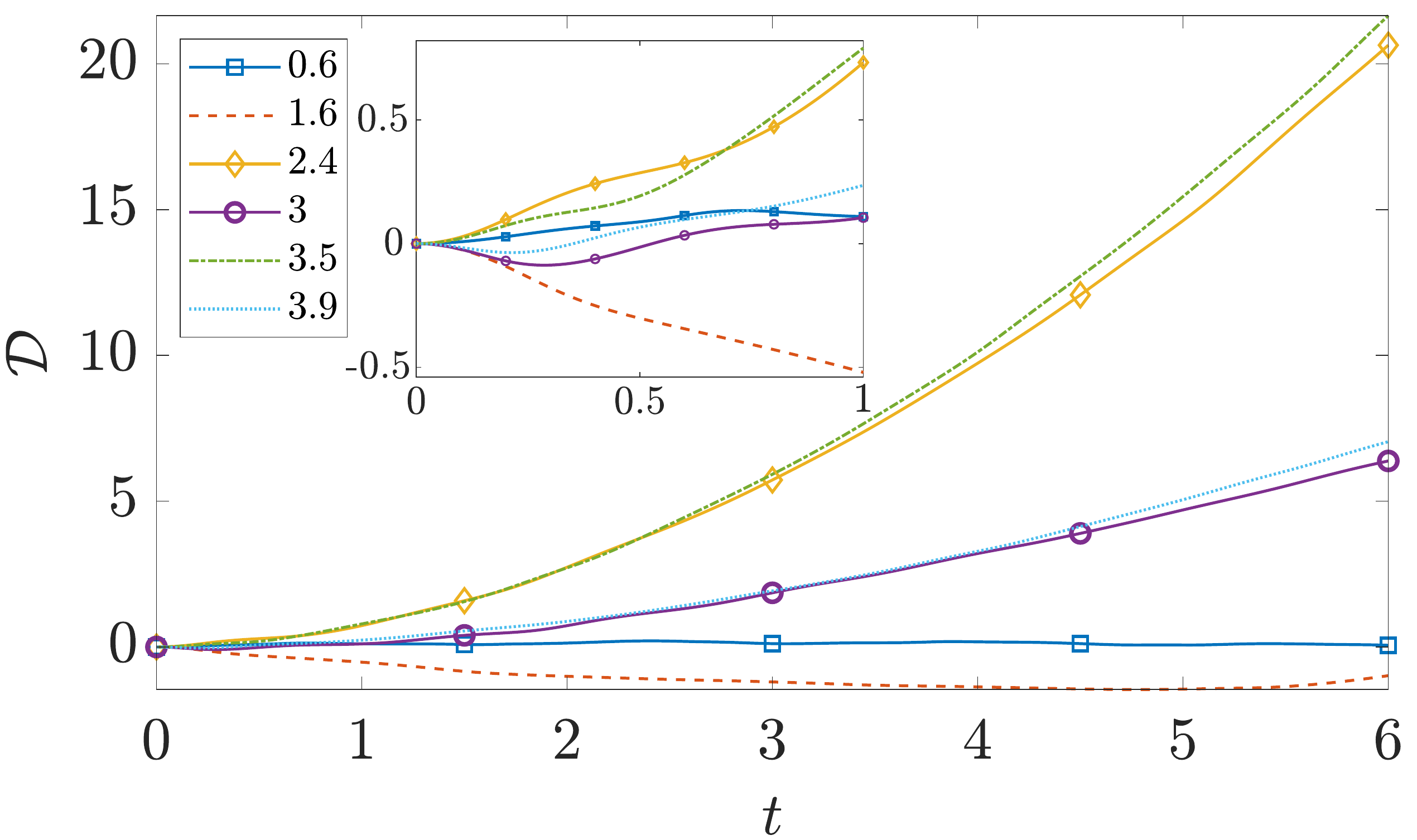}
	\caption{The COM displacement $\mathcal{D}$ varies with time for the evolutions starting from imbalanced two-Gaussian states in Eq.~\eqref{eq:13} with different initial separations $x_c=0.6$, 1.6, 2.4, 3, 3.5, and 3.9, respectively. Here, we have fixed $\eta=1/3$, $w=0.5$, and $\mathcal{C}_3 N^2=-20$. The sub-figure zooms in on the COM motion during the initial phase of the evolution.  }
	\label{fig7}
\end{figure}

In addition to the degree of asymmetry controlled by the relative amplitude ratio $\eta$, the self-acceleration behavior of the COM also depends on the initial separation $x_c$ between the two Gaussian wave packets. As observed in the variational results shown in Fig.~\ref{fig6}(e), the effective force acting on the COM alternates between positive and negative directions as the separation $x_c$ increases. To validate this behavior, we also examined the variation of the COM displacement during the dynamical evolution of the BEC, starting from the imbalanced two-Gaussian states by adjusting the initial separations. As illustrated in Fig.~\ref{fig7}, in the early stages of the evolution, the COM indeed moves in the positive or negative direction, depending on the initial separation. The results show a good agreement with the variational predictions, compared to those in Fig.~\ref{fig6}(e). It is important to emphasize that the variational prediction is valid only for the initial state. As time increases, the density distribution of the BEC would gradually deviate from the two-Gaussian profile [cf. Fig.~\ref{fig6}(a)], and as a result, the effective force experienced by the intermediate state may reverse its direction, causing the COM's movement to change direction. 

Furthermore, we would like to highlight that the self-acceleration behavior discussed above is based on the closed theoretical model described by Eq.~(\ref{eq:8}), which represents the dynamics of only the atoms. This model is derived by adiabatically eliminating the optical fields under the large detuning condition. From the perspective of this effective model, the atoms exhibit an unexpected self-acceleration behavior that seems to violate Newton's law of motion. However, the total momentum of the entire system, consisting of both atoms and optical fields, is conserved throughout the dynamical evolution, that is, Newton's law of motion holds for the entire system, including both atoms and optical fields~\cite{PhysRevX.5.011035}. To understand this, let us consider the transverse momenta of the driving fields. For the backward beams, the total transverse momentum flux before entering the BEC is
\begin{equation}
	\braket{\frac{d\mathbf{p}_{-}}{dt}}=-\sum_{\sigma=1,2}i\hbar\int \tilde{\mathcal{E}}^{-\ast}_\sigma(\mathbf{r}) \nabla_{\mathbf{r}} \tilde{\mathcal{E}}^-_\sigma(\mathbf{r})\mathrm{d}^2\mathbf{r} 
\end{equation}
with $\tilde{\mathcal{E}}^-_\sigma(\mathbf{r})=\sqrt{\frac{\epsilon_0}{2\hbar k_\sigma }}\mathcal{E}^-_\sigma(\mathbf{r})$. Similarly to the forward beams as described in Eq.~\eqref{eq:1}, the backward fields also acquire a phase shift after passing through the BEC as ${\mathcal{E}}^{-\prime}_\sigma(\mathbf{r})= {\mathcal{E}}^{-}_\sigma(\mathbf{r}) e^{-i\frac{k_{\sigma }\mu^2}{2\epsilon_0 \hbar \Delta^-_{\sigma }} {\rho}(\mathbf{r}) }$. 
The corresponding total transverse momentum flux of the backward beams after traversing the condensate reads 
\begin{equation}
	\begin{split}
		\braket{\frac{d\mathbf{p}_{-}'}{dt}}=& -\sum_{\sigma=1,2}i\hbar\int \tilde{\mathcal{E}}^{- \ast}_\sigma(\mathbf{r})\nabla_{\mathbf{r}}\tilde{\mathcal{E}}^-_\sigma(\mathbf{r})\mathrm{d}^2\mathbf{r}\\
		&-\sum_{\sigma=1,2}\frac{k_{\sigma }\mu^2}{2\epsilon_0 \Delta^-_{\sigma }}  \int \left|{\tilde{\mathcal{E}}}^{-}_\sigma(\mathbf{r})\right|^2 \nabla_{\mathbf{r}}\rho(\mathbf{r}) \mathrm{d}^2\mathbf{r}.
	\end{split}
\end{equation}
As a result, the momentum shift rate $d\Delta \mathbf{p}/dt=d\mathbf{p}_{-}'/dt-d\mathbf{p}_{-}/dt$ of the backward beams is given by
\begin{equation}
	\begin{split}
		\braket{\frac{d\Delta \mathbf{p}}{dt}}=& \sum_{\sigma=1,2}\frac{\mu^2}{4\hbar \Delta^-_{\sigma }} \int \rho(\mathbf{r})\nabla_{\mathbf{r}}\left|{{\mathcal{E}}}^{-}_\sigma(\mathbf{r})\right|^2\mathrm{d}^2\mathbf{r}\\
		=& \int \rho(\mathbf{r}) \nabla_{\mathbf{r}}V(\mathbf{r})\mathrm{d}^2\mathbf{r},
	\end{split}
	\label{eq:16}
\end{equation}
manifesting an effective force exerted on the optical fields. To derive the above equation, we have used the relation $\int\nabla_{\mathbf{r}}[\left|{{\mathcal{E}}}^{-}_\sigma(\mathbf{r})\right|^2\rho(\mathbf{r})]\mathrm{d}^2\mathbf{r}=0$. By repeating the same analysis for the forward beams, we find that the corresponding momentum shift flux vanishes. That is, only the backward beams eventually obtain a finite total transverse momentum. Compared with the force imposed on the atoms $\mathbf{F}=-\int \rho(\mathbf{r}) \nabla_{\mathbf{r}} V(\mathbf{r})\mathrm{d}^2\mathbf{r}$, one can observe that the forces acting on the light fields and the atoms are equal in magnitude but opposite in direction. This indicates a momentum transfer between the atoms and the optical fields, which ultimately leads to the motion of the COM of the BEC. 

Additionally, to verify that the self-acceleration behavior of the COM is solely caused by the nonreciprocal nature of the effective three-body interaction, we have also examined the dynamics of the condensate governed exclusively by the two-body interaction $V^{\rm II}$ as described in Eq.~(\ref{eq:6}) (not shown). In such a case where only the reciprocal two-body interaction is present, the COM remains stationary throughout the evolution. This contrasts sharply with the self-acceleration phenomenon driven by the three-body interaction, emphasizing the unique property of the non-reciprocity.

\section{Conclusions}
\label{sec:5}

In this work, we propose a scheme for generating effective three-body interactions in continuum quantum gases. We consider a light-atom coupling system composed of a quasi-2D BEC and two reflecting mirrors, with the BEC placed in front of the mirrors and illuminated by dichromatic light beams. In this configuration, the forward light acquires a phase shift depending on the density distribution of the atoms after it first passes through the BEC. The light then propagates towards the mirrors and is back-reflected. Subsequently, the backward light traverses the atoms for the second time, inducing feedback on the atoms. This feedback effect manifests as effective multi-body interactions between atoms under the condition of large detuning, where the optical fields can be adiabatically eliminated. Due to the compensation between the dichromatic lights, the three-body interaction is retained, while the two-body interaction can be entirely switched off. Based on recent experiments in similar setups~\cite{Robb2023PRR,PhysRevLett.132.143402}, this proposal can be implemented with current experimental techniques, as demonstrated in Refs.~\cite{PhysRevLett.121.073604,PhysRevA.103.023308}. In contrast to previous work, where three-body interactions typically emerge as higher-order contributions in the presence of dominant two-body effects, our proposal offers a highly flexible approach to study the effects driven purely by three-body interactions. This allows for the complete removal of the two-body effect while maintaining the three-body interaction.
   
Furthermore, this light-induced three-body interaction exhibits unique long-range and nonreciprocal properties, distinguishing it significantly from other synthetic three-body interactions studied previously. Unlike typical long-range interactions (e.g., Coulomb or dipole-dipole interactions), whose strength monotonically varies with the distance between particles, the long-range three-body interaction oscillates spatially, alternating between repulsion and attraction. Because of this unconventional behavior, the condensate can stabilize into emergent stationary states, including self-bound droplet clusters with various structures and a peculiar ring state. However, if the three-body interaction becomes sufficiently strong, the condensate loses stability and favors an unexpected \textit{diffusive collapse}. In addition to these self-bound stationary states, the nonreciprocal nature of the three-body interaction leads to a counterintuitive dynamical phenomenon: self-acceleration in the absence of an external force, seemingly violating Newton's law of motion.

Our proposal integrates several distinct aspects of particle-particle interactions, namely, long-range, multi-body, and nonreciprocal interactions, each of which plays a crucial role in many-body physics and has aroused significant attention in recent years. These are brought together within a simple light-atom coupling system. Given the high flexibility of this scheme and the remarkable properties of the effective interaction, it provides a promising avenue for exploring exotic physics related to multi-body interactions through the quantum simulation technique in cold atoms, e.g., statistical principles governed by nonreciprocal interactions~\cite{PhysRevX.12.010501,PhysRevX.5.011035}, self-bound droplets without quantum fluctuations~\cite{PhysRevLett.128.083401}, super-Heisenberg scaling measurement~\cite{PhysRevLett.105.180402,PhysRevLett.119.010403}.

\section*{Acknowledgement}
This work was supported by the National Nature Science Foundation of China (Grant No. 12104359, 11974334, and 12474366), Shaanxi Academy of Fundamental Sciences (Mathematics, Physics) (Grant No. 22JSY036), the National Key Research and Development Program of China (Grant No. 2021YFA1401700), and the Innovation Program for Quantum Science and Technology (Grant No. 2021ZD0301200 and 2024ZD0300600). Y.C.Z. acknowledges the support of the Xiaomi Young Talents program, Xi'an Jiaotong University through the ``Young Top Talents Support Plan'' and Basic Research Funding, as well as the High-performance Computing Platform of Xi'an Jiaotong University for computing facilities. H.P. acknowledges support from the NSF (Grant No. PHY-2207283) and the Welch Foundation (Grant No. C-1669).

\bibliography{mybib}
\end{document}